%% file: main.tex
\documentclass[sigconf]{acmart}

\usepackage[most]{tcolorbox}
\tcbuselibrary{listings,breakable}
\usepackage{amsmath,amsfonts}
\usepackage{algorithm}
\usepackage[caption=false,font=normalsize,labelfont=sf,textfont=sf]{subfig}
\usepackage{stfloats}
\usepackage{url}
\usepackage{verbatim}
\usepackage{amsmath,amssymb,amsfonts}
\usepackage{algorithmic}
\usepackage{caption}
\usepackage{textcomp}
\usepackage{xcolor}
\usepackage{soul}
\usepackage{array}
\usepackage{enumitem}
\usepackage{tcolorbox}
\usepackage{listings}
\usepackage{longtable}
\usepackage{tabularx}
\usepackage{booktabs} 
\usepackage{graphicx} 
\usepackage{multirow}
\usepackage{rotating}
\usepackage{booktabs} 
\usepackage{xcolor}
\usepackage{listings}
\usepackage{siunitx} 
\usepackage[utf8]{inputenc}
\usepackage{booktabs}
\usepackage{siunitx}
\usepackage{makecell}
\usepackage{caption}
\usepackage{float} 
\usepackage{xcolor,colortbl}
\usepackage{listings}
\usepackage{breakurl}
\usepackage{hyperref}
\usepackage{graphicx}
\usepackage{subcaption}
\AtBeginDocument{%
  }

\setcopyright{acmlicensed}
\copyrightyear{2018}
\acmYear{2018}
\acmDOI{XXXXXXX.XXXXXXX}

\acmJournal{POMACS}
\acmVolume{37}
\acmNumber{4}
\acmArticle{111}
\acmMonth{8}

\begin{document}

\title{A Deep Dive Into Large Language Model Code Generation Mistakes: What and Why?}

\author{QiHong Chen}
\email{chenqh@uci.edu}
\affiliation{%
  \institution{University of California, Irvine}
  \city{Irvine}
  \state{California}
  \country{USA}
}

\author{Jiachen Yu}
\email{jy91@illinois.edu}
\affiliation{%
  \institution{University of Illinois Urbana-Champaign}
  \city{Champaign and Urbana}
  \state{Illinois}
  \country{USA}
}

\author{Jiawei Li}
\email{jiawl28@uci.edu}
\affiliation{%
  \institution{University of California, Irvine}
  \city{Irvine}
  \state{California}
  \country{USA}
}

\author{Jiecheng Deng}
\email{jdeng043@ucr.edu}
\affiliation{%
  \institution{University of California, Riverside}
  \city{Riverside}
  \state{California}
  \country{USA}
}

\author{Justin Tian Jin Chen}
\email{chenjt3@uci.edu}
\affiliation{%
  \institution{University of California, Irvine}
  \city{Irvine}
  \state{California}
  \country{USA}
}

\author{Iftekhar Ahmed}
\email{iftekha@uci.edu}
\affiliation{%
  \institution{University of California, Irvine}
  \city{Irvine}
  \state{California}
  \country{USA}
}

\renewcommand{\shortauthors}{Trovato et al.}

\begin{abstract}
Recent advancements in Large Language Models (LLMs) have enabled their widespread use in automated code generation. However, these models still produce defective code that deviates from specifications. Previous research mainly focused on errors in LLMs with limited capabilities using small datasets, which restricts the comprehensiveness of identified mistakes. In this paper, we studied the mistakes made by two top-performing LLMs on an extensive dataset, identifying 17 types of non-syntactic mistakes, ten of which were overlooked by previous studies. We also identified six underlying reasons for these mistakes and proposed a method to verify the validity of these reasons. Furthermore, we explored the effectiveness of LLMs in detecting both mistakes and their causes. Our evaluation showed that GPT-4 has impressive performance when identifying mistakes in LLM-generated codes. Furthermore, GPT-4, using the ReAct prompting technique, achieved an F1 score of 0.78 in identifying reasons for mistakes. These findings provide valuable insights for improving the quality of LLM-generated code.
\end{abstract}
\maketitle





\input{01-intro}
\input{02-RW}

\input{03-setup}
\input{04-RQ1}

\input{05-RQ2}
\input{06-RQ3}

\input{07-discussion}
\input{08-ttv}

\input{09-conclusion}
\bibliographystyle{ACM-Reference-Format}
\bibliography{acmart}

\end{document}

%% file: 01-intro.tex
\section{Introduction}
\label{sec:intro}

Recent advancements in Large Language Models (LLMs) have garnered significant attention from both industry and academia, leading to a surge in LLM-based applications and research publications \cite{zamfiroiu2023chatgpt,fan2023automated, xu2022systematic, jain2022jigsaw, beaulieu2024evaluating}. These models have been trained with large data corpora that contain natural language texts and source code, showing high competency in following human instructions in a wide range of downstream tasks \cite{geng2024large,guo2024exploring,feng2024prompting}. One popular application of LLM is code generation, which utilizes LLMs to automatically generate source code for a given natural language requirement (i.e., coding question) \cite{zheng2023codegeex,mu2023clarifygpt,dong2023self,lin2024llm}. LLMs have shown impressive performance on this task~\cite{zhong2024ldb, hu2025qualityflow}. Thus, developers have begun integrating LLMs into their daily coding tasks in various stages of software development \cite{toth2024llms, castelberg2024llm,geng2024large,guo2024exploring}.

While LLMs are showing impressive performance, research indicates that they can produce factually incorrect outputs~\cite{chen2023hallucination, liu2024survey}. These mistakes pose a significant threat to the reliability of LLMs~\cite{liu2024exploring, rawte2023survey, chang2024survey}. For example, 
LLMs could recommend nonexistent third-party libraries to software developers, contributing to the distribution of software prone to exploitation \cite{mware}. Furthermore, companies such as Google have adopted LLMs to write codes for their software~\cite{google_code}. Consequently, mistakes in those automatically generated codes could harm the quality of the software. Therefore, uncovering and understanding the mistakes in LLM responses and their root causes has become urgent.

In recent years, the Software Engineering (SE) community has started investigating the mistakes in LLM-generated code and the underlying reasons. Fan et al.~\cite{fan2023automated} identified four categories of syntactic mistakes in LLM-generated codes, while Song et al.~\cite{song2023empirical} found seven categories of syntactic and non-syntactic mistakes. Although the mistakes identified in previous studies are valuable, several significant limitations remain. These studies focused exclusively on LLMs with limited capabilities—such as GPT-3.5, and Codex—potentially overlooking the mistakes of more advanced LLMs. In addition, their mistakes are collected from codes generated from relatively small datasets, including LeetCode and HumanEval, which limits the generalizability of their findings. Additionally, those works have focused exclusively on mistake in Python LLM-generated code, neglecting mistakes in other programming languages. However, identifying mistakes in LLM-generated code beyond Python is crucial, as these codes can also be employed in software development tasks that require different programming languages \cite{altherwi2021empirical}. 

To address these shortcomings, our study examines Python and Java code generated by top-performing LLMs using datasets that include nearly 11 times as many coding questions, enabling us to identify more errors than previous studies \cite{song2023empirical, fan2023automated, steenhoek2024err}. In addition to identifying additional mistakes, our study investigates the underlying reasons for the mistakes. This analysis is crucial, as it deepens our understanding of these mistakes and helps developers mitigate them when prompting LLMs to generate code. Finally, unlike previous studies \cite{steenhoek2024err} that relied solely on manual discussions to validate the reasons, we introduce a novel semi-automated verification method to substantiate that the identified reasons are indeed responsible for the mistakes, thereby enhancing the reliability of our findings. Specifically, our experiments aimed to address the following research questions:
\begin{itemize}
    \item RQ1: What mistakes do LLMs make in the code generation?
    \item RQ2: What are the underlying reasons for the LLM's mistakes in the code generation task?
    \item RQ3: How effectively can LLMs identify the mistakes and the underlying reasons in their generated code?
\end{itemize}

\noindent Specifically, this paper makes the following contributions:
\begin{enumerate}
\item A derived list of LLMs' code generation mistakes.
\item A derived list of reasons behind LLMs' mistakes in the code generation task.
\item A benchmark of 202 coding questions that LLMs could not solve correctly, along with the reasons.
\item An empirical investigation on LLM's ability to identify the mistakes and underlying reasons in its generated code.
\end{enumerate}



%% file: 02-RW.tex
\label{sec:rw}
\section{Related Works}
\subsection{LLMs for Code Generation}
Recent advancements in Natural Language Processing have led to the development of powerful LLMs, such as GPT-4 \cite{achiam2023gpt} and Claude 3 \cite{anthropic2024claude}, which have demonstrated exceptional performance in both general NLP tasks \cite{chang2024survey} and Software Engineering (SE) tasks like Code Generation. Code-specific LLMs (CodeLLMs), such as StarCoder \cite{lozhkov2024starcoder}, DeepSeek-Coder \cite{guo2024deepseek}, and Qwen2.5-Coder \cite{hui2024qwen2}, are trained primarily on source code corpora with objectives designed to capture syntactic and semantic code features. Thanks to their extensive training data, these models exhibit strong performance on downstream tasks through prompt engineering alone, without requiring fine-tuning. Recent studies have achieved state-of-the-art Code Generation results by leveraging LLM prompting techniques \cite{huang2023agentcoder,zhou2023language,zhong2024ldb,li2024acecoder}. Building on this research, we prompted LLMs to generate source code given problem specifications from widely used Code Generation benchmarks. We then analyzed these LLM-generated code to identify mistakes and their underlying causes.

\subsection{Mistakes in LLM Generated Code} 
Researchers have employed sound verifiers \cite{stechly2023gpt} to assess the correctness of LLM-generated code by enforcing strict rules or test cases. However, these verifiers are task-specific and require significant manual effort to design and implement. More recently, researchers have explored prompting LLMs to critique their own outputs, such as self-critique \cite{luo2023critique} and self-refinement \cite{madaan2024self}, in general NLP tasks. However, Code Generation differs fundamentally from these tasks as it requires understanding both natural language and source code. Fan et al. \cite{fan2023automated} found that Codex \cite{chen2021evaluating} makes errors similar to those of humans when solving LeetCode problems. Song et al. \cite{song2023empirical} identified syntactic and semantic mistakes in LLM-generated code on the HumanEval dataset. Zhang et al. \cite{zhang2024llm} manually analyzed Python3 CoderEval-generated code, categorizing eight types of mistakes, while Tambon et al. \cite{tambon2025bugs} examined 333 bugs from LLM-generated code on the same dataset, identifying ten mistake categories.

Despite these insights, existing studies have key limitations. First, they primarily analyze small datasets, limiting the comprehensiveness of identified mistakes. Second, they focus on earlier LLMs like ChatGPT and Codex, raising concerns about the relevance of their findings as newer, more capable models are introduced. Finally, they examine only Python3-generated code, overlooking potential variations in mistakes across programming languages. To address these gaps, we analyzed mistakes in LLM-generated Python and Java code across two datasets containing 2,268 coding questions, identifying both the errors and their underlying causes.

\subsection{Why LLMs Make Mistakes in Code Generation} 
Several studies have explored the underlying reasons for LLM-generated code mistakes. Jesse et al.\cite{jesse2023large} found that LLMs produce Simple and Stupid Bugs (SStuBs) due to quality issues in their training data. Kou et al.\cite{kou2024large} attributed errors to a misalignment between LLMs’ and human coders’ attention, analyzing mistakes through model attention patterns. Mu et al.~\cite{mu2023clarifygpt} reported that ambiguous or insufficient coding requirements contribute to errors and introduced ClarifyGPT, which prompts LLMs to ask clarifying questions to mitigate misunderstandings. In this study, we take the first step toward a comprehensive analysis of the reasons behind non-syntactic mistakes in LLM-generated code for both standalone and production-level functions. Unlike prior work, which examines mistakes from a single perspective, we consider all previously identified factors. Furthermore, we conduct experiments to verify that these factors directly contribute to the observed mistakes.

%% file: 03-setup.tex
\section{Experimental Setup}
\label{sec:setup}




\textbf{Non-syntactic Mistakes:} As LLMs grow larger, incorporating more parameters and extensive training data, they tend to produce fewer \textit{syntactic mistakes}—errors that prevent code from compiling— \cite{ma2023lms, sergeyuk2024using}. While syntactic mistakes are easily detected through compilation, \textit{non-syntactic mistakes} are more elusive, manifesting only in specific test cases or remaining undetected if test coverage is insufficient. We argue that \textit{non-syntactic mistakes} reveal not only limitations in LLMs’ programming capabilities but also flaws in their reasoning and understanding abilities. Addressing these issues could inform the development of more reliable LLMs for code generation. Therefore, this study focuses on \textit{non-syntactic mistakes}, which fall into two categories: (1) the generated code compiles but encounters runtime errors (i.e., crashes during execution) and (2) the generated code compiles and runs but fails to meet the functional requirements specified, resulting in test failures.

\noindent \textbf{Code Generation Datasets:} \label{sec:datasets} We selected two widely used multilingual datasets, namely HumanEval-X~\cite{zheng2023codegeex} and MBXP~\cite{athiwaratkun2022multi}. The HumanEval-X dataset comprises 328 coding problems—164 in Python and 164 in Java—with an average of seven test cases per coding question. In contrast, the MBXP dataset includes 1,940 coding coding questions, with 974 in python3 and 966 in java, and an average of three test cases per coding question. We selected datasets supporting python3 and java due to their extensive use in industry and open-source projects \cite{tiobe}.

\noindent \textbf{Model Selection:} To ensure the generalizability of our results, we selected two LLMs with cutting-edge performance in Code Generation:
\noindent\textbf{GPT-4}: GPT-4 is one of the most powerful general-purpose LLMs developed by OpenAI. It has demonstrated exceptional performance in numerous SE tasks including Code Generation. We adopt GPT-4-0125-preview (GPT-4)~\cite{achiam2023gpt} in our experiments. 
\noindent\textbf{Qwen2.5-Coder}: Qwen2.5-Coder developed by Alibaba Group is a CodeLLM series with different model sizes, which have shown state-of-the-art performance in a wide range of SE tasks such as Code Generation. In our study, we select qwen2.5-coder-14b-instruct \cite{hui2024qwen2}.



\noindent \textbf{Data Preparation:}\label{sec:data_preparation} To collect LLM-generated code snippets with \textit{non-syntactic mistakes}, we prompted GPT-4 and Qwen2.5-Coder to solve coding questions from the selected datasets (Section \ref{sec:datasets}). Following best practices~\cite{he2024does,hu2025qualityflow}, we designed our prompts to include task instructions, coding question specifications, function signatures, and input-output examples demonstrating the required functionality. To ensure reproducibility, we set the LLMs' temperature to 0 and executed multiple runs for each coding problem to obtain the most deterministic outputs~\cite{peng2023towards}.

Next, we executed all test cases associated with the coding questions (detailed performance metrics for each LLM on these datasets, measured by Pass@1 score, are provided in our replication package~\cite{replication}). We then filtered out code snippets that failed to compile due to \textit{syntactic mistakes} and those that passed all test cases (i.e., correct solutions). In total, we discarded 27 LLM-generated code snippets with syntactic mistakes, corresponding to 25 coding questions. We collected their actual outputs and retained their test failure information for the remaining code snippets that failed at least one test case or encountered runtime errors (\textit{non-syntactic mistakes}).

%% file: 04-RQ1.tex
\section{RQ1:What mistakes do LLMs make in the code generation?}
\label{sec:rq1} 


\subsection{Methodology to Answer RQ1}
\noindent\textbf{Obtain Correct Code:}\label{sec:obtain_correct_code} 
To identify non-syntactic mistakes in LLM-generated code, we compared the incorrect code to the correct solution. However, we observed that LLM-generated code may differ from the ground truth in terms of the algorithms and data structures used. This is reflected in the low Jaccard similarity scores (0.24-0.28 for Qwen2.5-Coder, and 0.22-0.44 for GPT-4) between the LLM-generated code and the ground truth. Discarding all the incorrect code with such divergences would significantly reduce the number of samples in our dataset, limiting the comprehensiveness and effectiveness of our analysis. To overcome this issue, we applied Automated Program Repair (APR) techniques to correct the generated code\cite{le2012systematic,xia2023keep}. APR makes small adjustments to the incorrect code while preserving the original algorithmic approach. This technique allows us to compare LLM-generated code with APR-corrected code, providing a more accurate understanding of the LLM’s mistakes.

We utilized CHATREPAIR~\cite{xia2023keep}, a state-of-the-art APR technique, which fixes the code through an iterative conversational process. The tool interacts with the LLM based on error messages and test failure information, refining the code in multiple iterations. The number of trials and iterations per trial can be adjusted as hyperparameters, and we used the default settings as recommended by Xia et al.~\cite{xia2023keep} If CHATREPAIR could not fix the code within 30 trials, we resorted to using the ground truth from the dataset.

To validate the effectiveness of the APR process, we measured the Jaccard similarity between the APR-repaired code and the original incorrect code. We found a high similarity (0.88 for Qwen2.5-Coder, and 0.72-0.86 for GPT-4), indicating that the APR repairs maintained the original algorithmic approach. This suggests that our APR technique was effective in preserving the intended solution structure while correcting the mistakes\cite{weissman2015identifying,li2024coir}. The effectiveness of CHATREPAIR is further detailed in our replication package\cite{replication}. We compared Jaccard similarities for each LLM-generated code with non-syntactic mistakes. If the APR-fixed code had a higher similarity to the LLM code than the ground truth, we used the APR-fixed code as the reference. In 99\% of cases, the APR-fixed code showed higher similarity to the LLM code than the ground truth.

\noindent\textbf{Mistake Identification:}\label{sec:mistake_identification} To identify mistakes, we followed a two-step process. First, each author independently reviewed the incorrect LLM-generated code and its correct version, focusing on critical control statements (e.g., 'if' and 'while') that affected program logic. We also analyzed test case failures, variable values during execution, and runtime logs. Each author created an overview of the mistakes, although some mistakes might have been missed. In the second step, the authors (each with at least five years of Python and Java programming experience) met to discuss and validate the mistakes, proposing valid inputs to trigger each identified issue. This process resulted in a comprehensive list of mistakes for each incorrect LLM-generated code.


\noindent\textbf{Mistake Category Identification:}\label{sec:mistake_category_identification} After identifying mistakes in LLM-generated code, all four authors conducted open coding \cite{khandkar2009open} using a negotiated agreement process \cite{hansen2002recommendations} to categorize them. We began by considering the likely process an LLM follows when solving a coding problem—interpreting the specification, devising a solution, and implementing the code. Since these steps involve reasoning, algorithm selection, and programming skills (e.g., data structures, operators, APIs), each author independently analyzed the mistakes and proposed categories based on these aspects.

Next, we met to discuss and align our categorizations. For each mistake, all authors presented their views, followed by a group discussion focusing on core elements such as algorithms and code constructs. If disagreements remained, a fifth author reviewed the arguments and made the final decision. After reaching a consensus, we examined the categories to merge semantically similar ones, with the fifth author resolving any unresolved debates. Finally, we refined category names to ensure clarity and informativeness.

\noindent\textbf{Mistake Severity Identification:} After identifying errors in LLM-generated code, we assessed the severity of each mistake based on its impact on functional correctness. This analysis offers two key benefits: (1) it helps practitioners prioritize debugging efforts by focusing on critical issues, and (2) it serves as an indicator of code quality, enabling more objective comparisons of LLM performance.

Each author reviewed the incorrect code to understand its overall structure, focusing on key elements such as parameters, outputs, and comments to infer intended functionality. We analyzed the function in segments, starting with variable initialization and data structures, followed by control flows (e.g., conditionals, loops) and exception handling to examine how the code processes different inputs. Combining these observations, each author formed a detailed understanding of the adopted algorithm and its functionality, which was then compared against the coding question’s specification to assess deviations. Based on this gap, each author independently categorized the severity of the algorithmic mistakes.

Following the individual analyses, all four authors met to discuss and agree on the severity of each incorrect code. Each author presented their assessment, including the identified algorithm and its implemented functionality, and the group compared these against the intended functionality. If disagreements remained, a fifth author was consulted to make the final determination. After reaching a consensus, we reviewed and merged semantically similar categories, with the fifth author resolving any remaining disputes. Lastly, we refined category names to ensure clarity and informativeness.

\begin{table*}[htbp]
\centering
\scriptsize 
\caption{Mistake Categories, Mistake Types and Severity Frequency (Highlighted Types Are The Newly Identified Mistakes)}
\label{tab:mistakes}
\resizebox{\textwidth}{!}{%
  \begin{tabular}{l p{0.50\columnwidth} c c c}
    \toprule
    \textbf{Mistake Categories} & \textbf{Mistake Types} & \textbf{Flawed (FADE)} & \textbf{Partial (PADE)} & \textbf{Divergent (DADE)} \\
    \midrule
    \multirow{2}{*}{Condition Error} 
      & Conditional Misalignment Error (15.65\%) & \multirow{2}{*}{45.55\%} & \multirow{2}{*}{49.74\%} & \multirow{2}{*}{4.71\%} \\
      & Missing Condition Error (14.24\%)        &                     &                     &                     \\
    \midrule
    \multirow{3}{*}{Index Issue} 
      & \cellcolor{yellow}Indexing Convention Mismatch Error (2.50\%)         & \multirow{3}{*}{48.57\%} & \multirow{3}{*}{37.14\%} & \multirow{3}{*}{14.29\%} \\
      & \cellcolor{yellow}Index Utilization Error (0.78\%)          &                     &                     &                     \\
      &  Miscalculated Index Error (2.19\%)  &                     &                     &                     \\
    \midrule
    \multirow{3}{*}{Return Issue} 
      & \cellcolor{yellow}Missing Return Error (1.88\%)              & \multirow{3}{*}{95.83\%} & \multirow{3}{*}{4.17\%}  & \multirow{3}{*}{0\%} \\
      & \cellcolor{yellow}Minor Output Format Mismatch Error (8.92\%) &                     &                     &                     \\
      & \cellcolor{yellow}Wrong Variable Return (0.47\%)             &                     &                     &                     \\
    \midrule
    Math Knowledge Issue 
      & \cellcolor{yellow}Incorrect Math Knowledge Error (14.24\%) & 7.69\%  & 20.88\% & 71.43\% \\
    \midrule
    \multirow{3}{*}{API Usage} 
      & \cellcolor{yellow}API Return Handling Error (1.10\%)           & \multirow{3}{*}{40.82\%} & \multirow{3}{*}{55.10\%} & \multirow{3}{*}{4.08\%} \\
      & \cellcolor{yellow}Inappropriate Data Structure (3.44\%)&                     &                     &                     \\
      &  API Misuse Error (10.80\%)   &                     &                     &                     \\
      
    \midrule
    \multirow{2}{*}{Erroneous Pattern} 
      & \cellcolor{yellow}Incorrect Mask Error (1.56\%)  & \multirow{2}{*}{15\%}  & \multirow{2}{*}{80\%} & \multirow{2}{*}{5\%} \\
      & \cellcolor{yellow}Incorrect Regex Error (1.56\%) &                     &                     &                     \\
    \midrule
    \multirow{3}{*}{Incorrect Functionality} 
      & Flawed Implementation (1.72\%)           & \multirow{3}{*}{8.33\%} & \multirow{3}{*}{44.70\%} & \multirow{3}{*}{46.97} \\
      & Partial Implementation (9.23\%)         &                     &                     &                     \\
      & Completely Incorrect Implementation (9.70\%) &                 &                     &                     \\
    \bottomrule
  \end{tabular}
}%
\end{table*}


\begin{figure*}[htbp]
    \centering
    \begin{minipage}{0.35\textwidth}
         \centering
         \includegraphics[width=\textwidth]{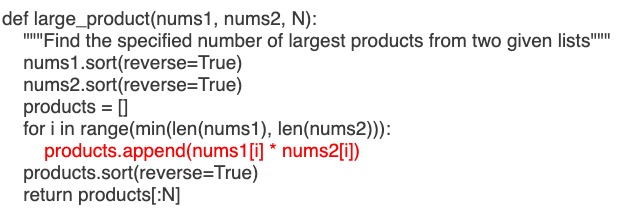}
         \\[0.5ex]
         \textbf{(a)} Index Utilization Error Example
         \label{pic:IUE}
    \end{minipage}
    \hfill
    \begin{minipage}{0.35\textwidth}
         \centering
         \includegraphics[width=\textwidth]{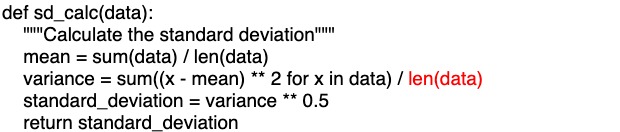}
         \\[0.5ex]
         \textbf{(b)} Incorrect Math Knowledge Error Example
         \label{pic:IMKE}
    \end{minipage}
    
    \vskip\baselineskip 
    
    \begin{minipage}{0.35\textwidth}
         \centering
         \includegraphics[width=\textwidth]{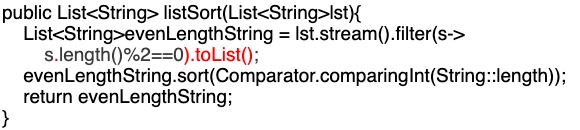}
         \\[0.5ex]
         \textbf{(c)} API Return Handling Error Example
         \label{pic:ARHE}
    \end{minipage}
    \hfill
    \begin{minipage}{0.35\textwidth}
         \centering
         \includegraphics[width=\textwidth]{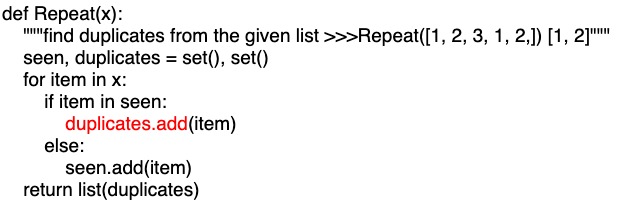}
         \\[0.5ex]
         \textbf{(d)} Inappropriate Data Structure Example
         \label{pic:ARHE}
    \end{minipage}    
    \caption{LLM Code Generation Mistakes Examples-IUE, IMKE, ARHE, IDS}
    \label{fig:combined_mistake_examples1}
\end{figure*}

\begin{figure}[htbp]
    \centering
    \begin{minipage}{0.45\textwidth}
         \centering
         \includegraphics[width=\textwidth]{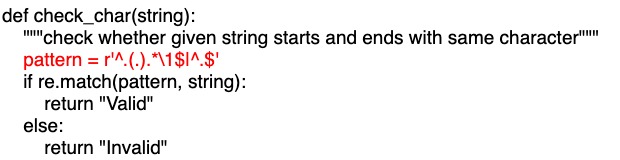}
         \\[0.5ex]
         \textbf{(a)} Incorrect Regex Error Example
         \label{pic:IRE}
    \end{minipage}
    \hfill
    \begin{minipage}{0.45\textwidth}
         \centering
         \includegraphics[width=\textwidth]{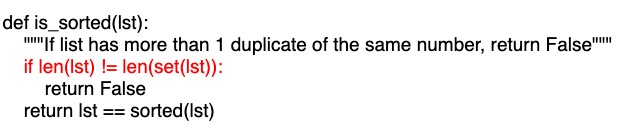}
         \\[0.5ex]
         \textbf{(b)} Conditional Misalignment Error Example
         \label{pic:CME}
    \end{minipage}
    \caption{LLM Code Generation Mistakes Examples-IRE, CME}
    \label{fig:combined_mistake_examples2}
\end{figure}



\subsection{Answer to RQ1: Non-Syntactic Mistake Analysis} 
The overall mistake categories are summarized in Table~\ref{tab:mistakes}. Through manual inspection, we identified seven high-level mistake categories: \textit{Condition Error, Index Issue, Return Issue, Math Knowledge Issue, API Usage, Erroneous Pattern, and Incorrect Functionality}, which are further divided into 17 specific mistake types. Due to space constraints, we present examples for only a subset of these mistakes; complete examples are available in our replication package~\cite{replication}. Among the 17 mistake types, ten types were overlooked by prior studies~\cite{song2023empirical, zhang2024llm, tambon2025bugs} (highlighted in Table \ref{tab:mistakes}). The oversight can be explained by two main factors. First, our analysis not only considered smaller datasets such as HumanEval but also included LLM-generated code from the MBXP dataset, which comprises 1,948 coding questions—nearly 11 times the 164 questions in the HumanEval dataset. This more diverse dataset choice enabled us to explore a broader range of coding scenarios, thereby uncovering additional mistakes. Second, by analyzing LLM-generated code in both Python and java, our study was able to identify mistakes that may be unique to specific programming languages. Below, we provide definitions and examples of these ten non-syntactic mistake types:


\noindent\textbf{Indexing Convention Mismatch Error (ICME) (2.50\%)}: arises when the LLM misaligns its assumed index system (often 0-based) with the index system specified. This could lead to LLM returning the wrong index or makes out of bound access.

\noindent\textbf{Index Utilization Error (IUE) (0.78\%)}: arises when the LLM misuses indices (not due to index arithmetic calculation or index convention system) to access the search space—either under-utilizing them, which leads to exploring only a subset of the necessary candidate solutions, or over-utilizing them, by examining a broader range than required. In both cases, improper index usage overlooks valid solutions or expends unnecessary computational effort, ultimately leading to suboptimal or incorrect outcomes. An example is shown in Figure~\ref{fig:combined_mistake_examples1}(a). The coding question requires the LLM to identify the specified number of largest products from two given lists. To do so, it must first compute the products for all pairs formed from the two lists. However, the LLM only considers pairs with matching indices (marked in red), thereby excluding other combinations that might yield larger products.

\noindent\textbf{Missing Return Error (MRE) (1.88\%)}: arises when the LLM correctly implements the desired functionality but does not have the return statement to return the result in the code.

\noindent\textbf{Minor Output Format Mismatch Error (MOFME) (8.92\%)}: arises when the generated code is functionally correct but the output deviates from the coding question only in terms of its representation. This may involve differences in data type, numeric precision, or other formatting details, resulting in a mismatch between the expected and actual output format.

\noindent\textbf{Wrong Variable Return (WVR) (0.47\%)}: arises when the generated code is functionally correct, and the result is stored in a designated variable. However, instead of returning this variable at the end of the code, the incorrect code returns a different one, leading to an incorrect output.

\noindent\textbf{Incorrect Math Knowledge Error (IMKE) (14.24\%)}: arises when the LLM uses the incorrect mathematical concepts, formulas, or properties to solve the coding question. Consequently, the generated code deviates from the intended mathematical reasoning required to solve the problem. It can manifest in multiple ways, such as improperly sequencing operations (e.g., performing addition before multiplication, contrary to the standard order of operations) or using the wrong formula to calculate the math sequences. An example is shown in Figure \ref{fig:combined_mistake_examples1}(b). The coding question requires the LLM to calculate the standard deviation. The LLM-generated code follows the standard approach by computing the mean and variance. However, the code incorrectly calculates the variance by dividing len(data) instead of len(data)-1 (marked in red).

\noindent\textbf{API Return Handling Error (ARHE) (1.10\%)}: arises when the output of an API call is mishandled due to incorrect assumptions about its data type or mutability. In essence, the mistake involves treating the returned value of an API call as if it were of a different type or had different properties than it actually had. An example is shown in Figure \ref{fig:combined_mistake_examples1}(c), an LLM incorrectly uses a \textit{stream} call and then attempts to convert it to a list object (in red). The \textit{stream} call returns a non-mutable type, so the subsequent \textit{toList} function call fails, highlighting the LLM's misuse of the \textit{toList} method on the \textit{stream's} return object.

\noindent\textbf{Inappropriate Data Structure (IDS) (3.44\%)}: arises when the LLM does not adequately account for the intrinsic properties of the adopted data structure, such as automatic duplicate removal or the absence of order. An example is shown in Figure \ref{fig:combined_mistake_examples1}(d). The coding question requires the LLM to identify duplicate integers from a list while preserving their original order. To accomplish this, the LLM uses a set to store duplicates (marked in red). Although a set effectively eliminates redundant entries, its inherent lack of order results in the loss of the original sequence of duplicates.

\noindent\textbf{Incorrect Mask Error (IME) (1.56\%)}: arises when the LLM uses the incorrect mask to perform the bit operation.

\noindent\textbf{Incorrect Regex Error (IRE) (1.56\%)}: arises when the LLM uses an incorrect regex pattern to match the given string. An example is shown in Figure~\ref{fig:combined_mistake_examples2}(a). The coding question specifies the LLM to check whether the given string starts and ends with the same character using regex. However, the LLM adopted a regex pattern (marked in red) that checks for the second character and the last character instead.

Besides those ten mistakes types, we have also identified seven mistake types that align with prior studies \cite{song2023empirical, zhang2024llm, tambon2025bugs}. Below, we describe those seven types:

\noindent\textbf{Conditional Misalignment Error (CME) (15.65\%)}: 
arises when the LLM generates a condition that misaligns with the coding question, resulting in flawed conditional logic. This mistake is not limited to if or else-if statements but also includes conditions in loops such as while and for. For example, as shown in Figure~\ref{fig:combined_mistake_examples2}(b), the coding question requires returning false only if a number appears three or more times in the list. However, the LLM-generated code incorrectly returns false when a number appears just twice, as seen in the faulty condition if len(lst) != len(set(lst)) (highlighted in red).


\noindent\textbf{Missing Condition Error (MCE) (14.24\%)}: arises when the LLM-generated code omits a crucial condition outlined in the coding question specification or needed to address flaws in the approach, such as handling specific input scenarios or corner cases. Unlike the \textit{Conditional Misalignment Error}, where the condition deviates from the specification, the Missing Condition Error arises when the required condition is entirely absent.


\noindent\textbf{Miscalculated Index Error (MIE) (2.19\%)}: arises when the generated code performs an incorrect index calculation, leading to accessing an uninitialized or incorrect index. Unlike an \textit{Indexing Convention Mismatch Error}, where the model uses a different indexing system than specified, in this case the LLM adheres to the correct indexing convention but miscalculates the index.

\noindent\textbf{API Misuse Error (AME) (10.80\%)}: 
arises when the LLM uses the wrong API call or incorrectly invokes the correct API with the wrong parameters or incorrect number of arguments. For example, a coding question specifies that for a given number \textit{a} and a digit value \textit{digits}, the LLM should round \textit{a}  to the specified digits. However, the LLM mistakenly uses Python's round function with the expression \textit{round(a + 10**(-len(str(a))-1), digits)}. In Python, the round function does not always round up, depending on the specified digit value. Since the question requires the number to always round up, the LLM should have used the ceil function instead.


\noindent\textbf{Flawed Implementation (FI) (1.72\%)}: arises when the adopted algorithm in the generated code is mostly correct with only some minor logical error and its implemented functionality mostly aligns with required functionality.

\noindent\textbf{Partial Implementation (PI) (9.23\%)}: arises when the adopted algorithm is on the right track but with noticeable flaws that need to be addressed. 

\noindent\textbf{Completely Incorrect Implementation (CII) (9.70\%)}: arises when the LLM-generated code is completely incorrect. Specifically, the functionality implemented in the LLM-generated code completely differs from the desired functionality described in the coding question specification.


\subsection{Mistake Severity Categorization}
This subsection presents the identified severity levels.

\noindent\textbf{Flawed Algorithm Design Error (FADE)}: arises when the LLM's approach is generally on the right track—showing reasonable reasoning—but the algorithm contains flaws. These flaws may allow the code to produce correct or nearly correct outputs for some inputs, yet it fails on others due to overlooked corner cases or inherent logical issues that cause incorrect results even when all cases are handled.

\noindent\textbf{Partial Algorithm Design Error (PADE)}: arises when the LLM’s approach has a significant gap but remains aligned with the goal of solving the coding question. However, part of the required functionality—clearly specified in the problem description or examples—is either completely missing or incorrectly implemented.

\noindent\textbf{Divergent Algorithm Design Error (DADE)}: arises when the LLM’s implementation is entirely off track, substantially deviating from the intended functionality described in the coding question specification and examples. This includes cases where the code performs the opposite of what is asked, attempts to solve an unrelated problem, or overlooks most or all required functionalities.

\subsection{Analysis of Mistake Types and Severity}
We present the number of LLM-generated codes associated with each mistake category and the frequency of their severity levels in the last three columns of Table~\ref{tab:mistakes}.

Our analysis shows that mistake types like condition errors, return issues, API misuse, and erroneous patterns are primarily tied to flawed or incomplete algorithm designs. In contrast, math knowledge mistakes often stem from divergent algorithm designs, highlighting that even top-performing LLMs frequently misapply mathematical concepts when solving coding tasks—consistent with our observation that LLMs struggle with core mathematical definitions and sequences.

Additionally, we found that 37\% to 40\% of all mistakes are related to flawed or partial algorithm designs, suggesting that LLMs generally understand the coding questions and attempt reasonable approaches. However, minor output format mismatches are also common (detailed in our replication package~\cite{replication}), indicating that while LLMs grasp functional requirements, they often miss formatting details. Such oversights, though seemingly minor, can lead to cascading failures in systems relying on the generated code. Therefore, practitioners should pay special attention to verifying output formats when using LLM-generated code.




%% file: 05-RQ2.tex
\section{RQ2: What are the underlying reasons for the LLM's mistakes in the code generation task?}
\label{sec:rq2}


\subsection{Methodology to Answer RQ2}

\subsubsection{Reason Identification}\label{sec: core_reasons_discussion} 

To identify the reasons behind non-syntactic mistakes, we analyzed the typical process LLMs follow when solving coding problems: interpreting the coding question, devising a logical strategy, and implementing the code. We hypothesized that four factors influence this process: \textit{(i) Phrasing used in the prompt}, \textit{(ii) Presentation structure}. Additionally, input-output demonstrations help the LLM confirm its understanding and verify its approach \cite{chen2024premise, gao2023makes}, leading to our third hypothesis: \textit{iii) Input-output pair demonstration quality}. Finally, when generating code, the LLM relies on programming skills acquired during training \cite{jesse2023large}. This suggests a fourth contributing factor: \textit{iv) Training-Induced mistakes}. To identify the underlying causes of mistakes, we employed an open coding process \cite{khandkar2009open}  combined with a negotiated agreement approach\cite{hansen2002recommendations}. Four authors independently reviewed incorrect LLM-generated code, reference code, and coding questions to determine the reasons for errors, followed by group discussions to resolve disagreements. Once validated, the reasons were categorized and empirically tested by modifying coding questions based on the identified factors and prompting the LLM to regenerate the code. The reason was considered validated if the regenerated code passed all test cases. Below, we detail the analysis performed for the four hypothesized factors.

\textit{i) Phrasing Used In The Prompt:} The LLM's comprehension of the coding question specification plays a crucial role in the quality of its generated code. One key factor influencing this comprehension is the phrasing and word choices used in the question. Ambiguities in the coding question can lead to misunderstandings, which manifest in three primary ways: \textbf{Ambiguous Wording:} The question may be phrased in a way that allows multiple interpretations, leading to inconsistencies in the LLM's response. \textbf{Misalignment with Function Names:} The intended meaning of the question may not fully align with the provided function name, potentially causing misinterpretation of the task. \textbf{Misleading Parameter Names:} Parameter names in the function signature may inadvertently suggest meanings that do not exist. For example, consider a function intended to calculate the area of a parallelogram, with the signature:\textit{def calculate\_area(b, h)}. While the variable names b and h might suggest ``base'' and ``height,'' they may actually represent two arbitrary sides of the parallelogram. This discrepancy can mislead the LLM, affecting its interpretation and leading to incorrect implementations.


To identify mistakes introduced due to this hypothesis, all authors independently reviewed the referenced correct code, coding question specifications, and identifiers of functions and parameters in the function signature. The analysis followed a structured four-step process:
\textbf{Identifying the requirements of the coding question} – Each author independently determined what the coding question was asking for.
\textbf{Understanding the LLM’s mistake:} Authors examined what the LLM actually did when it made the mistake.
\textbf{Group discussion:} The team discussed discrepancies between the intended requirements and the LLM's actual implementation, particularly focusing on whether the phrasing in the prompt contributed to the misunderstanding.
\textbf{Generating paraphrased versions:} Sentences containing potentially ambiguous phrases were paraphrased using GPT-4, with explicit instructions to preserve their original semantic meaning.

GPT-4 was chosen for paraphrasing due to its strong natural language understanding capabilities \cite{karanikolas2023large}. The temperature was set to 0.5, ensuring a balance between diversity and controlled randomness, allowing creative yet semantically consistent variations. To validate the paraphrased sentences, all authors reviewed the original and modified versions to confirm that the intended meaning remained unchanged. If most authors found that a paraphrase altered the meaning, it was excluded from the analysis.
\textit{ii) Presentation Structure:}\label{sec: question_present} 
LLMs are built on transformer architectures that use attention mechanisms to represent the focus. Insufficient attention allocation can cause LLMs to overlook certain content. To address this issue, it is crucial to enhance the LLM's attention to the overlooked sections of the specification. We hypothesized that the position of information within the input influences the LLM’s attention distribution. Thus, repositioning relevant information—while preserving its semantic meaning—may encourage the LLM to allocate more attention to it. It is important to note that the \textit{presentation structure} emphasizes the position of the overlooked specification in the prompt, whereas the \textit{phrasing used in the prompt} centers on the specific wording of the specification.

To identify mistakes introduced due to this hypothesis, each author individually and then through negotiated agreement \cite{hansen2002recommendations} reached a consensus on the \textit{missing functionality}. Then, we moved the description of the missing functionality to a before and after location in the coding question (see Figure \ref{fig:combined_reason_examples}(d) as an example). To ensure this adjustment did not change the meaning of the coding question, all authors read and compared the descriptions before and after the change. In cases where changing the position of the word phrases would affect the meaning of the coding question, we instead added the phrase ``note that'' immediately before the missing functionality in the original description to force the LLM to pay more attention.


\textit{iii) Input-output pair demonstration quality:} Besides the specification, the input-output pair demonstrations also help the LLM understand the intended functionality and the input/output format. Consequently, non-syntactic mistakes made by LLMs can often be traced back to the quality of these demonstrations—for example, how thoroughly they convey the desired functionality, specify the output format, or address potential corner cases.

We first collected all LLM-generated codes that successfully passed the demonstration examples but failed on other test cases to identify mistakes arising from this factor. We then applied program slicing techniques \cite{xu2005brief, binkley1996program, zhong2024ldb} to break down the incorrect code into smaller segments. For each segment, we traced intermediate variable values using program tracing \cite{larus1993efficient} and analyzed the code logic to understand why it produced correct results for demonstration examples but failed under different test conditions. Each author independently conducted this analysis and, through negotiated agreement \cite{hansen2002recommendations}, reached a consensus on the LLM’s \textit{reasoning and thinking logic about the given input-output pairs}. The group then compared this reasoning with the original coding question to determine whether the LLM had misinterpreted the problem or overfitted to the provided demonstrations.


To verify the reason, we employed two strategies. First, we augmented the coding question with additional examples to improve the LLM's understanding of the task. These examples were generated by prompting GPT-4 to design new test inputs, after which we executed the ground truth code to obtain the corresponding expected outputs. We selected GPT-4 for test case design due to its strong code understanding ability \cite{achiam2023gpt}. To ensure the quality of these demonstrations, we discarded any inputs for which the ground truth code failed to produce an output. Second, we revised the prompt to instruct the LLM to consider not only the provided inputs but also other valid inputs.



\textit{iv) Training-Induced:} LLMs acquire programming skills by learning frequent patterns in their training data. However, this reliance on common patterns can lead to mistakes when subtle, language-specific features are not well captured during training. These features include operators (e.g., negation, inversion) and API calls to built-in or third-party libraries. A key issue arises when an LLM confuses the functionality of a method or operator across different programming languages. For example, the function call \textit{round(-5.5)} is syntactically correct in both Python and Java, and in most cases, their outputs align. However, for certain inputs—such as -5.5—the results differ: Python rounds it to -6, while Java rounds it to -5. Such errors likely stem from limitations in training data, either due to insufficient exposure to these nuances or an inadequate number of relevant examples. As a result, when generating code, the LLM may inadvertently apply the semantics of one language to another, leading to non-syntactic mistakes.

To investigate whether LLMs make non-syntactic mistakes due to failing to recognize subtle differences between identically named functions and operators across programming languages, we first compiled a list of functions and operators from Python and Java documentation that share the same name but differ in functionality \cite{pydoc, javadoc}, return types, or signatures. We then identified coding questions where the LLM's implementation was correct in one language (e.g., Python) but incorrect in the other (e.g., Java). Using AST, we pinpointed lines containing function calls and operators in both implementations and excluded those not on our list. We analyzed the LLM’s use of functions, including parameter types, number of arguments, and return value handling, along with its use of operators. When the LLM invoked functions or used operators similarly across both languages—despite subtle differences—it suggested that the LLM either overlooked or was unaware of these distinctions. Furthermore, if the LLM processed a function’s return value as identical across languages, despite differences, this indicated an incorrect understanding of the function's return details.

\subsection{Answer to RQ2: Reasons of Non-syntactic Mistakes}
\label{sec:result_rq2}

In this section, we report the six reasons identified. 

\begin{figure*}[htbp]
    \centering
    \begin{minipage}{0.45\textwidth}
         \centering
         \includegraphics[width=\textwidth]{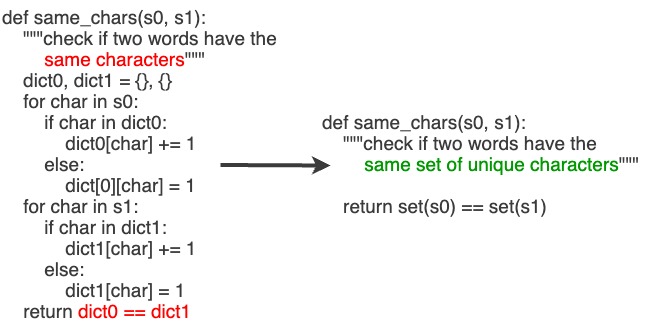}
         \\[0.5ex]
         \textbf{(a)} Misleading Coding Question Specification Example
         \label{pic:MCQS}
    \end{minipage}
    \hfill
    \begin{minipage}{0.45\textwidth}
         \centering
         \includegraphics[width=\textwidth]{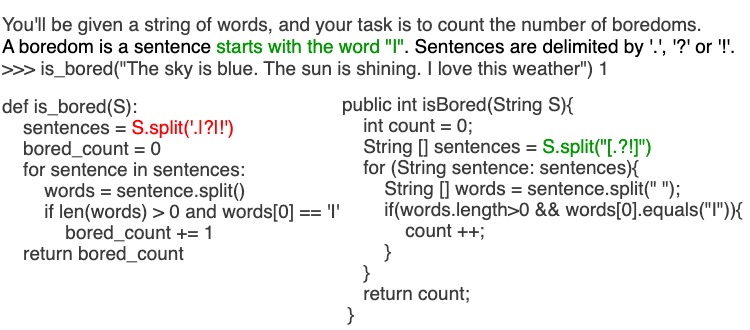}
         \\[0.5ex]
         \textbf{(b)} Incorrect Trained Knowledge Example
         \label{pic:EC}
    \end{minipage}
    
    \vskip\baselineskip 
    
    \begin{minipage}{0.45\textwidth}
         \centering
         \includegraphics[width=\textwidth]{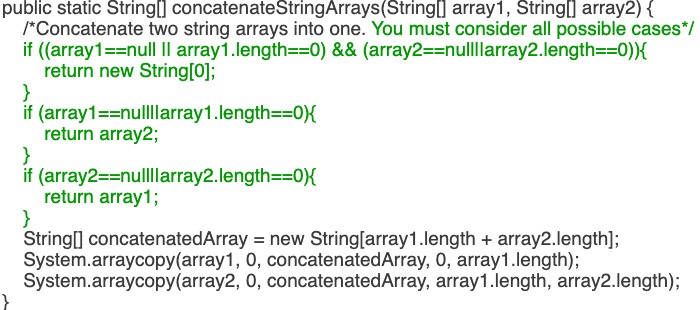}
         \\[0.5ex]
         \textbf{(c)} Edge Case Example
         \label{pic:PS}
    \end{minipage}
    \hfill
    \begin{minipage}{0.45\textwidth}
         \centering
         \includegraphics[width=\textwidth]{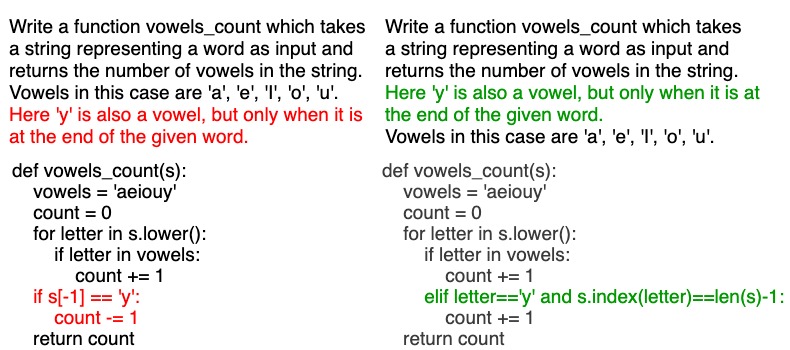}
         \\[0.5ex]
         \textbf{(d)} Positional Sensitivity Example
         \label{pic:ITK}
    \end{minipage}
    
    \caption{Example Reasons for LLM Code Generation Mistakes}
    \label{fig:combined_reason_examples}
\end{figure*}


\noindent\textbf{1. Misleading Coding Question Specification (MCQS) (56.19\%)}\label{sec: mislead_result} refers to the cases where LLMs are confused by specific phrases used in the coding question specification. We found that such ambiguities can lead to a range of mistakes, including condition errors, index issues, missing return errors, and minor output format mismatches. For example, consider Figure~\ref{fig:combined_reason_examples}(a), where the LLM was given the following coding question specification: \textit{``Check if two words have the same characters."} Upon analyzing the incorrect code generated by the LLM, we observed that it used two dictionaries to separately count the frequency of characters in each word. This suggested that the LLM interpreted the question as checking whether the two words were anagrams—that is, whether they contained the same characters with the same frequencies. Recognizing this misinterpretation, we hypothesized that the confusion arose from the phrase \textit{``same characters"}, as the word \textit{``same"} may have led the LLM to consider character frequencies rather than just character presence. To verify our hypothesis, we prompted GPT-4 to rephrase the instruction to reduce ambiguity. It generated the revised phrase: \textit{``same set of unique characters.''}. We then asked the LLM to regenerate the code using the revised question specification:\textit{``Check if two words have the same set of unique characters.''} The newly generated code passed all test cases, confirming our hypothesis that the phrase \textit{``same characters''} introduced ambiguity, leading to the LLM’s misinterpretation.


\noindent\textbf{2. Poor Input-output Demonstration (PIOD) (21.26\%)} 
occurs when low-quality or insufficient input-output examples lead to LLM misunderstandings, affecting code correctness. Ambiguous or limited demonstrations can result in multiple interpretations, causing the LLM to generate flawed yet seemingly correct outputs. This aligns with Tambon et al.'s findings on LLM bias towards provided examples \cite{tambon2025bugs}. Additionally, when coding questions lack sufficient examples or explicit output format specifications, LLM-generated code often exhibits minor formatting errors, likely due to the model's inability to infer the expected return format.

\noindent\textbf{3. Incorrect Trained Knowledge (ITK) (5.71\%)} 
refers to cases where LLMs generate incorrect code due to flawed knowledge learned during training, supporting the \textit{Training-Induced} hypothesis. Specifically, LLMs sometimes mislearn details of language-specific features like APIs, libraries, and operators, leading to errors such as API misuse, return handling issues, and indexing mismatches. An example of LLM misusing the API call is shown in Figure~\ref{fig:combined_reason_examples}(b). In this example, the LLM generated one Python code (left) and one Java code (right) for the same coding question. However, the Python code mistakenly uses the \textit{split} function since LLM incorrectly assumed that the \textit{split} function in Python is the same as that in Java, which can split the string by regular expression. This issue reflects the limitations of LLM training on vast, unlabeled datasets, where the lack of language context in training examples can induce persistent misconceptions, a ``necessary evil''~\cite{radford2019language}. 

\noindent\textbf{4. Misleading Function Signature (MFS) (4.44\%)} 
refers to cases where LLMs are misled by function names in the coding question specification that conflict with the intended functionality, aligning with the \textit{phrasing used in the prompt} hypothesis. When function names suggest a different task than described, LLMs may generate incorrect code, causing errors like condition mistakes, index issues, or output mismatches. For instance, in a task requiring the removal of all elements that appear more than once, the function name \textit{remove\_duplicates} led GPT-4 to mistakenly remove only duplicate instances. Renaming it to \textit{removeALLElementsOccurMoreThanOnce} resulted in correct code, confirming that misleading function signatures can impair LLM performance \cite{yang2022important}.

\noindent\textbf{5. Edge Case (EC) (4.86\%)} occurs when LLMs fail to account for all possible corner cases. This relates to the \textit{Input-output pair demonstration quality} hypothesis because LLMs could be biased by the input-output pair demonstration and overlook the possible edge cases.  As a result, LLM-generated code may work for provided examples but fail on edge cases, leading to errors like conditional misalignment or missing conditions. For instance, as shown in Figure~\ref{fig:combined_reason_examples}(c), GPT-4’s code failed when an input array was null. However, adding the phrase \textit{``You must consider all possible cases"} to the prompt led to correctly handling null and empty inputs.

\noindent\textbf{6. Positional Sensitivity (PS) (4.12\%)} 
refers to cases where LLMs overlook or misinterpret parts of a coding question due to lack of attention\cite{chen2024premise}, supporting the \textit{Presentation structure} hypothesis. This often occurs in longer prompts, leading to partial implementations or missing condition errors. For example, shown in Figure~\ref{fig:combined_reason_examples}(d), when asked to count vowels and treat \text{`y'} as a vowel only if it appears at the end (as shown in the left figure, marked in red), the LLM misapplied this rule. By simply repositioning the instruction about \text{`y'} to an earlier part of the prompt (as shown in the right figure, marked in green), the LLM produced correct code, indicating that non-syntactic mistakes can result from positional sensitivity to how information is presented.

%% file: 06-RQ3.tex
\section{RQ3: How effectively can LLMs identify the mistakes and the underlying reasons in thier generated code?}\label{sec:rq3}


In the first two research questions, we manually analyzed mistakes in LLM-generated code and identified their underlying causes, but this approach was time-consuming and impractical for everyday use. Given the strong reasoning abilities demonstrated by recent LLM advancements \cite{kojima2022large, gilardi2023chatgpt}, we explore their potential for automating the identification of mistakes and uncovering their causes. 



\subsection{Methodology to Answer RQ3}

\subsubsection{Mistake Identification} To evaluate LLM's abilities in identifying the mistakes in the generated code, we selected GPT-4 because of its widespread adoption by researchers~\cite{luo2023critique} for critiquing the LLM-generated responses to various reasoning-intensive tasks and its strong code understanding ability \cite{achiam2023gpt}.
 
\noindent\textbf{Prompt Construction:} We provided GPT-4 with the coding question, incorrect LLM-generated code, and test case failure information. The failure information included the test code, error message, expected output, and actual output of the LLM-generated code. We then instructed GPT-4 to identify potential non-syntactic mistakes (the full prompt is included in our replication package~\cite{replication}). Afterward, we manually reviewed GPT-4's responses and compared them to the mistakes identified in RQ1 (Section \ref{sec:rq1}). For responses that did not align with our identified mistakes, we verified them using a procedure similar to that in Section \ref{sec:rq1}.


\noindent\textbf{Evaluation Metric:} To evaluate GPT-4's performance using precision and Coverage Rate (CR). Precision was calculated based on true positives (actual mistakes reported by the LLM) and false positives (hallucinated mistakes). The coverage rate (CR) is defined as:
\begin{equation}
\label{eqn:recall}
\text{CR} = \frac{\text{\# Identified True Mistakes}}{\text{\# Total Mistakes}}
\end{equation}
where Identified True Mistakes represent the actual mistakes reported by LLM, while Total Mistakes are true mistakes identified by either humans or LLM. 

\subsubsection{Reason Identification} To evaluate LLM's abilities in identifying the reasons for LLM's mistakes, we first created an evaluation benchmark consisting of 202 instances, where each instance consists of incorrect LLM-generated code, its corresponding coding question and the human-labeled reasons. We created this benchmark based on the reasons identified in Section \ref{sec:rq2}. Subsequently, we experimented with three approaches using GPT-4: \textit{Base Prompt}, \textit{Advanced Prompt}, and \textit{Advanced Prompt+ReAct}. The prompts are included in our replication package \cite{replication}. Below, we describe each prompt.

(i) \textbf{Base Prompt:} \label{sec:base_prompt} 
As a baseline, we prompted GPT-4 to identify the reasons behind non-syntactic mistakes in the incorrect code. The prompt included the coding question specification, the incorrect LLM-generated code, its actual output, and the test failure information, along with instruction to prompt the LLM to leverage the provided information to analyze how LLM was misled into generating the incorrect code and to provide a comprehensive explanation. Importantly, we did not include human-labeled reasons or their definitions in the prompt. This approach allowed us to assess whether the LLMs could independently identify reasons comparable to those identified by humans.

(ii) \textbf{Advanced Prompt:}\label{sec:advanced_prompt} 
In addition to the information in the  \textit{Base Prompt}, this prompt included definitions of all six human-labeled reasons (Section \ref{sec:result_rq2}). We expected these definitions to help GPT-4 identify reasons with the intended granularity. The LLMs were instructed to report the human-identified reasons and provide explanations along with their reasoning steps.


(iii) \textbf{Advanced Prompt+ReAct:} \label{sec:react_prompt} 
We employed the ReAct~\cite{yao2022react} (Reasoning + Actions) prompting technique, which enables LLMs to actively reason while retrieving relevant information through various actions. This approach allows GPT-4 to dynamically select and apply retrieved information, enhancing their reasoning process. Given its effectiveness in tasks requiring extensive context for informed reasoning and generation~\cite{yao2022react}, ReAct is well-suited for our reason identification task. To guide GPT-4 in identifying underlying reasons, we developed three distinct tools. Each tool replicates the manual steps used to pinpoint these reasons (if applicable) or provides necessary information for GPT-4 to reason.



\noindent \textbf{Description of the tools implemented:}


(ii) \textbf{Function Call Analysis Tool:} 
The function call analysis tool extracts and analyzes third-party library and built-in function invocations from incorrect LLM-generated code, retrieving their corresponding official documentation along with the code. This helps identify incorrect function usage by enabling LLMs to compare the invoked functions with their expected behavior. We parsed Python and Java code using the \textit{ast}\cite{ast} and \textit{javalang}\cite{javalang} modules, respectively. For Python, we retrieved documentation using the \textit{render\_doc} function from the \textit{pydoc}\cite{pydoc} library. For Java, we accessed documentation from the Oracle API\cite{oracle}, using \textit{BeautifulSoup}~\cite{bs4} to parse and extract relevant information. Our rationale for designing this tool is to mimic our manual process to provide the LLM with the called API documentation information.


(iii) \textbf{Function Signature Explainer Tool:} 
This tool is designed to identify the misleading function signature by automating the manual steps we took in Section\ref{sec:result_rq2}. In specific, this tool clarifies the intended functionality of a function based on its signature. It extracts the function signature and parameter names from the coding question and prompts GPT-4 to generate a summary describing the function's purpose. GPT-4 subsequently leverages this summary to understand the functionality described by the function signature and compares it with the coding question specification to assess the clarity of the function signature.

(iv) \textbf{Coding Question Specification Ambiguity Check Tool:} 
This tool evaluates the ambiguity of a coding question specification using the self-consistency method \cite{wang2203self}. The premise is that if an LLM has a consistent understanding of a question, repeated queries should yield similar responses. To implement this, we provide GPT-4 with the coding question and a human-written summary. GPT-4 then generates five summaries of the question, each with a temperature setting of 0.5 for balanced randomness and consistency. We convert these summaries into vector embeddings using the \textit{all-mpnet-base-v2} model \cite{stmodels}, which captures semantic meaning \cite{reimers2019sentence}. We compute pairwise cosine similarity scores among these embeddings. If most similarity scores are below 0.8 \cite{zhou2022problems}, we classify the coding question as ambiguous.

\noindent\textbf{Evaluation:} To evaluate our approaches, all authors independently reviewed the LLM-generated reasons and their explanations, comparing them to our evaluation benchmark to determine semantic equivalence with human-labeled reasons. Any disagreements were resolved through a consensus-driven discussion using the negotiated agreement approach \cite{hansen2002recommendations}. A reason was considered correct if it matched the corresponding reason in our benchmark.
\begin{table}[htbp]
\centering
\footnotesize
\caption{GPT-4 Reason Identification F1 Score}
\label{tab:rq4_table_f1}
\begin{tabularx}{\columnwidth}{l *{3}{>{\centering\arraybackslash}X}}
\toprule
\textbf{Reason} 
  & \textbf{Base Prompt} 
  & \textbf{Advanced Prompt} 
  & \textbf{Advanced ReAct} \\
\midrule
Misleading Coding Question Specification    & 0.87 & 0.91 & 0.93 \\
Poor Input Output Demonstration             & 0.67 & 0.72 & 0.87 \\
Edge Case                                   & 0.78 & 0.81 & 0.83 \\
Misleading Function Signature               & 0.62 & 0.63 & 0.68 \\
Positional Sensitivity                      & 0.25 & 0.62 & 0.65 \\
Incorrect Trained Knowledge                 & 0.64 & 0.68 & 0.70 \\
Average                                     & 0.64 & 0.73 & 0.78 \\
\bottomrule
\end{tabularx}
\end{table}


\subsection{Answer to RQ3: Performance of Mistake and Reason Identification} 

\subsubsection{Mistake Identification Result} Overall, GPT-4 demonstrates performance comparable to human evaluators on both HumanEval-X and MBXP. Specifically, GPT-4 achieved a precision of 0.97 on HumanEval-X and 0.95 on MBXP, with a Coverage Rate (CR) of 0.94 on HumanEval-X and MBXP datasets. In contrast, human evaluators reached a precision of 1 on both datasets, along with a CR of 0.98 on HumanEval-X and 0.99 on MBXP. These results suggest LLMs like GPT-4 are effective at identifying mistakes in LLM-generated code, including non-syntactic mistakes that humans may overlook. 

\subsubsection{Reason Identification Result} The performance of GPT-4 is shown in Table~\ref{tab:rq4_table_f1}. Due to the space constraint, we only reported the F1 score, and keep results in other metrics in our replication package\cite{replication}. For the \textit{Base Prompt}, GPT-4 achieved an average F1 score of 0.64, with an impressive performance in identifying misleading coding question specification. Using the \textit{Advanced Prompt}, the F1 score increased to 0.73. Compared to \textit{Base Prompt}, the \textit{Advanced Prompt} improved the performance in identifying positional sensitivity (from 0.25 to 0.62) and incorrect training knowledge (from 0.64 to 0.68). We posit that these improvements were largely due to the inclusion of human-identified reasons as they guided the LLMs’ focus toward previously overlooked aspects. In addition, the \textit{Advanced Prompt+ReAct} approach showed further improvement, with the F1 score rising from 0.73 to 0.78 with further improvement in identifying misleading coding question specification (from 0.91 to 0.93). These results indicate that top-performing LLMs can successfully identify most of the reasons, with the exceptions of issues related to positional sensitivity. Future work should focus on developing improved tools and methodologies to enhance LLMs’ ability to detect these shortcomings.

%% file: 08-ttv.tex
\section{Threats to Validity}

In this study, we have taken measures to mitigate potential threats affecting our findings' validity. Nevertheless, it is possible that some of our measures might not have been effective.

\textbf{Construct validity} The quality of the code generated by LLMs may have been influenced by our prompt design. While we followed best practices for prompt creation and in-context examples~\cite{gao2023makes,ibm,awesomeprompt}, we did not employ all possible advanced prompting techniques. Additionally, we relied on test cases from the datasets to assess the correctness of the generated and APR-repaired code. However, these test cases might not be comprehensive~\cite{kochhar2015code}. To address these concerns, we used widely adopted code generation benchmarks that are adopted by existing code generation studies.

\textbf{Internal validity} Our experiment relies heavily on manual examination, which may introduce limitations such as incorrect understanding and overlooked mistakes or reasons. To mitigate these risks, we adhered to established practices such as open coding and negotiated agreement~\cite{fereday2006demonstrating}, involving multiple authors in thorough discussions to reach a consensus on the findings.

\textbf{External validity} Our findings may not be generalizable to LLM-generated code from other benchmarks or programming languages beyond Java and Python and all available LLMs. However, since GPT-4 and Qwen2.5-Coder come from different organizations with potentially different training data and model parameters, their mistakes could offer insights applicable to a broader range of LLMs. Moreover, the identified reasons for the mistakes in LLM-generated code are not exhaustive since, with the evolution of LLMs, their mistakes and  underlying reasons also evolve. 

%% file: 09-conclusion.tex
\section{Conclusion}
\label{conclusion}
In this work, we identified 17 types of non-syntactic mistakes, and six reasons behind these mistakes in LLM-generated code. 
Moreover, we explored LLM's abilities to automatically identify the mistakes and underlying reasons. Overall, our findings are valuable for practitioners using LLM-generated code and for researchers developing new LLM-based code generation techniques. Our replication package is available at~\cite{replication}.